\documentclass[showpacs,showkeys,floatfix]{revtex4}
\usepackage[dvips]{graphicx}
\usepackage{epsfig}

\usepackage[cp1251]{inputenc}
\usepackage[english]{babel}

\begin{document}

\title{Revivals in Caldeira - Leggett 
Hamiltonian Dynamics.}
 
\author{V.A.Benderskii} 
\affiliation {Institute of Problems of Chemical Physics, RAS \\ 142432 Moscow
Region, Chernogolovka, Russia} 

\author{E. I. Kats}  
\affiliation{L. D. Landau Institute for Theoretical Physics, RAS, Moscow, Russia}

\author{A.S. Kotkin} 
\affiliation {Institute of Problems of Chemical Physics, RAS \\ 142432 Moscow
Region, Chernogolovka, Russia} 

\date{\today}

\begin{abstract}
In this work we reconsider the problem discussed by Caldeira and Leggett (CL),  A.O.Caldeira and A.J.Leggett, Ann. Phys., ${\bf 149}$, 374 (1983), 
and  generalize their results for a quantum oscillator $s$ coupled bilinearly to a reservoir $R$ 
with dense discrete spectrum of harmonic oscillators with close to $\omega _s$ frequencies. 
We show that for such systems dynamic evolution of any state of the CL Hamiltonian 
consists
of recurrence cycles with partial revivals of the initial state amplitude.
This revival or Loschmidt echo appears in each cycle due to reverse $R$ to $s$ transitions
(not existing in the limit of a macroscopic bath). 
Width and number of the Loschmidt echo components increase with the recurrence cycle number $k$, eventually leading to
irregular, stochastic-like time evolution.
Standard for continuous spectrum thermal bath CL dynamics dissipative dynamics (exponential decay) takes place only within the initial cycle $k=0$. 
In terms of the effective CL Hamiltonian, where the reservoir degrees of freedom are integrated out,
our results mean oscillatory (time periodic) behavior of the effective mass. Only in the initial
cycle, we found the known for CL Hamiltonian exponential in time increase of the effective mass.  
We do believe that the basic ideas inspiring our work can be applied to a large variety
of interesting 
nano-systems and molecular complexes.

\end{abstract}


\maketitle

{\bf{Introduction.}}
${\, }$ In this work we are dealing with dynamics of 
nano-particles or relatively large  molecular complexes.
In this case the system has a large (of the order of $10^2\, -\, 10^4$) number of internal vibration modes with a dense discrete spectrum.
Characteristic mean interlevel spacing ${\bar {\Delta }_0} \simeq 1\, - \,100 \, cm ^{-1}$,
in high-frequency region ($1000 \, -\, 2000\, cm^{-1}$)
corresponds to Poincare recurrence periods in the range $0.1 \, - \, 10\, ps$ routinely probed by modern optical methods.
Such dynamics of the initial excited state coupled to the reservoir of vibration states is in the heart
of the realm of radiationless transitions \cite{SC94}, \cite{1}, \cite{2}, \cite{2a}.
These experiments show that in many cases characteristic life time of lowest excited states, is on the
order of recurrence cycle period. Therefore the both phenomena (namely, decay of the state $s$ and exchange with $R$ states)
has to be treated together.
Although all robust dynamic characteristics of radiationless transitions are described
by the CL Hamiltonian \cite{CL83}, \cite{LC87} which reads (in the self-evident notations with $p_s$ and $q_s$ momentum
and coordinate of the harmonic oscillator coupled  bilinearly (with coupling $C_n$) to the discrete states of the reservoir,
also formed by harmonic oscillators $p_n$ and $q_n$)
\begin{eqnarray}
\label{bk1}
H = \frac{1}{2}(p_s^2 + \omega _s^2 q_s^2) + \frac{1}{2} \sum _{n} (p_n^2 + \omega _n^2 q_n^2) + q_s\sum _{n} C_n q_n
\, 
\end{eqnarray}
the Hamiltonian (\ref{bk1}) has been investigated with all details only in the limit of continuous spectrum
thermal bath. In this limit the stored in the excited state energy is
exponentially in time dissipated via Ohmic friction. Therefore many relevant for nano-systems and relatively large molecular
complexes dynamic aspects of revivals of the initial states are merely disregarded.
Surprisingly for us, scanning the literature we did not find any paper treating theoretically discrete spectrum CL model.

Our model is applicable to a standard setup used for radiationless transitions, and includes the following main ingredients. 
Preparation of the initial excited state $s$, coupled strongly (matrix elements $C_n$) 
to other high frequency vibration states $\{n\}$. These states
form reservoir $R$ with a dense discrete spectrum of levels distributed around
the excited initial state, i.e., resonance region width is smaller than characteristic frequencies
\begin{eqnarray}
\label{bk2}
|\omega - \omega _s| \ll \omega _s\, ,\, \omega _n
\, .
\end{eqnarray}
In the continuous spectrum reservoir one can integrate out the reservoir degrees of freedom and then end up
with the following effective Hamiltonian $H_{eff}^c$ bearing dissipative physics
\cite{YS94}
\begin{eqnarray}
\label{bk3}
H_{eff}^c = \exp (-\eta t)\frac{1}{2}P_s^2 + \exp (+\eta t) \frac{1}{2}Q_s^2 
\, ,
\end{eqnarray}
where entering effective Hamiltonian 
canonical momentum and coordinate
$P_s$ and $Q_s$ (satisfying the standard commutation relations), and ''friction'' coefficient $\eta $
characterizes the coupling to the reservoir, and it determines the time dependent effective mass.
Eq. (\ref{bk3}) is a Hamiltonian of damped oscillator under external force determined by initial values of the
reservoir momenta and coordinates.
Irreversible transition to the reservoir states yields to exponential decay of the initial state, and that is all
what one can expect from full CL (\ref{bk1}) (or its reduced version (\ref{bk3})) Hamiltonian for the continuous
thermal bath. The aim of our work is to see how this conclusion above (and the effective Hamiltonian (\ref{bk3}))
has to be modified for systems where the vibrational state reservoir is characterized by discrete
spectrum. 

{\bf{Secular equation for CL Hamiltonian with discrete spectrum reservoir.}}
${\, }$  
The equations of motion for the Hamiltonian of bilinearly coupled oscillators (\ref{bk1}) 
\begin{eqnarray}
\label{bk3a}
{\ddot {q}}_s + \omega _s^2 = - \sum _{n} C_n q_n \, ;\, {\ddot {q}}_n + \omega _n^2 q_n = - C_n q_s 
\, ,
\end{eqnarray}
supplemented by the initial conditions
\begin{eqnarray}
\label{bk3b}
q_s(0) = q_{s0}\, ,\, {\dot {q}}_s (0) = {\dot {q}}_{s0} \, , \, q_n(0) = q_{n0}\, , \, {\dot {q}}_{n0}
\, 
\end{eqnarray}
can be easily solved by Laplace transformation \cite{LC87}, \cite{BF07}
\begin{eqnarray}
\label{bk4}
{\tilde {q}}_{s, \{n\}}(u) = \int _{0}^{\infty } q_{s, \{n\}}(t)\, \exp (u t)\, dt 
\, .
\end{eqnarray}
Then 
the secular equation to find eigen-frequencies
\begin{eqnarray}
\label{bk7}
F(\omega ) = \omega ^2 - \omega _s^2  - \sum _{n} \frac{C_n^2}{\omega ^2 -  \omega _n^2} = 0
\, .
\end{eqnarray}
In the resonance conditions (\ref{bk2}), Eq. (\ref{bk7}) reads as
\begin{eqnarray}
\label{bk8}
F(\epsilon ) = F_0(\epsilon ) + \epsilon F_1(\epsilon )
\, ,
\end{eqnarray}
where $\epsilon \equiv \omega - \omega _s$ and
\begin{eqnarray}
\label{bk9}
F_0(\epsilon ) = 2\omega _s\left (\epsilon - \sum _{n} \frac{B_n^2}{\epsilon -  \epsilon _n}
\right )\, ;\, F_1(\epsilon ) = \epsilon + \sum _{n} \frac{B_n^2}{\epsilon -  \epsilon _n}
\left (1 + \frac{\epsilon _n}{2\omega _s}\right )^{-1}\, ,
\end{eqnarray}
with the renormalized coupling constants
\begin{eqnarray}
\label{bk10}
B_n = \frac{C_n}{2\omega _s}\left (1 + \frac{\epsilon _n}{2\omega _s}\right )^{-1/2}\, .
\end{eqnarray}

{\bf{Zwanzig approximation.}}
${\, }$ 
To simplify the calculations we start with Zwanzig model \cite{ZW60} assuming
equidistant in the resonance region the reservoir spectrum $\omega _n \equiv \omega _s + n$ and $B_n \equiv B$.
Then the secular equation (\ref{bk9}) acquires the compact form
\begin{eqnarray}
\label{bk12}
F_0(\epsilon ) = \epsilon - \eta \cot (\pi \epsilon ) 
, ,
\end{eqnarray}
where the friction coefficient $\eta = \pi B^2$ is the full rate of the transitions from the $s$ oscillator into all $R$ states
In terms of the harmonic oscillator coupled to a discrete spectrum reservoir,
its behavior is dramatically different from that for the continuous spectrum thermal bath.
In the latter case the friction coefficient $\eta $ is a time independent constant. Therefore
the coupling yields to irreversible decay (dissipation of the initially excited state). It is not the
case for the discrete spectrum. The effective ''friction coefficient'' 
is time dependent (e.g., it equals zero in a certain time within each of the recurrence cycle), and it leads
to Loschmidt echo phenomenon and non-trivial quantum dynamics.

{\bf{CL effective Hamiltonian for discrete spectrum reservoir.}}
${\, }$  
Having in hands the CL Hamiltonian and the equations of motion we can express the amplitude
of the $s$ oscillator as a series over the initial values with time dependent coefficients 
\begin{eqnarray}
\label{bk15}
q_s(t) = a_1(t)q_s(0) + a_2(t){\dot {q}}_s (0) + \sum _{n}(a_{n1}(t)q_n(0) + a_{n2}(t){\dot{q}}_{n}(0))
\, ,
\end{eqnarray}
and similarly for the Laplace transforms
\begin{eqnarray}
\label{bk16}
{\tilde {q}}_s(u) = {\tilde {a}}_1(u)q_s(0) + {\tilde {a}}_2(u){\dot {q}}_s (0) + \sum _{n}({\tilde {a}}_{n1}(u)q_n(0) + {\tilde {a}}_{n2}(u){\dot{q}}_{n}(0))
\, .
\end{eqnarray}
This expansion (\ref{bk15}) is determined by the poles of the function $1/F_0(\epsilon)$ which are the roots of the secular
equation. Namely the time dependent coefficients are
\begin{eqnarray}
\label{bk18}
a_1(t) = \left (\cos (\omega _s t) + \frac{\eta }{2\omega _s}\sin (\omega _st) \right ) W(t)
\, ;\, a_2(t) = \frac{\sin (\omega _s t)}{\omega _s}W(t) 
\, .
\end{eqnarray}
Slow (in a scale $\omega _s^{-1}$) variations of the amplitude of $s$ oscillator is described by the function 
$W$, which is represented in a Fourier series over the Hamiltonian eigen frequencies 
\begin{eqnarray}
\label{bk19a}
W(t) = \sum _{n \geq 0}\frac{2 \eta \cos (\epsilon _n t)}{(\eta ^2 (1 + 1/(\pi \eta )) + \epsilon _n^2)}
\, .
\end{eqnarray}
The series is calculated by Poisson summation formula and then it is reduced to a sum over partial recurrence 
cycle amplitudes \cite{BF07}, \cite{BK08}, \cite{BK09}
\begin{eqnarray}
\label{bk19}
W(t) = \sum _{k=0}^{[t/2\pi ]}W_k (\tau _k)\, ; \, W_k(\tau _k) = -\frac{\tau _k}{k}\exp (-\tau _k/2)\theta (\tau _k) L_{k-1}^1(\tau _k)
\, ,
\end{eqnarray}
where, $[x]$ means the integer part of $x$, $\theta (x)$ is the Heaviside step-function, $L_{k-1}^1$ is adjoint Laguerre polynomial \cite{BE53}, and the local recurrence 
cycle ''time'' is defined as $\tau _k = \eta (t -2 k\pi )$.
These expressions (\ref{bk15}) - (\ref{bk19}) manifest that time evolution of the excited
initial state coupled to the discrete spectrum reservoir can be reduced to one dimensional damped
oscillator with time dependent friction. Only for the initial $k=0$ cycle, when from (\ref{bk19})
$W_0(t) = \exp (-\eta t/2)$ we get simple exponential decay as it is the case for CL dynamics with
the continuous spectrum thermal bath.
From (\ref{bk15}), (\ref{bk19}) we conclude that time evolution of $q_s$ and $p_s$ is determined by the functions
$a_1(t)$ and $a_2(t)$ which depend on the oscillator frequency $\omega _s$ and friction coefficient $\eta $
but independent of time evolution of the reservoir states in a phase space $\{ q_n\, ,\, p_n\}$ orthogonal to the
phase space $q_s\, , \, p_s$.
Then it is convenient to make a projection \cite{LC87}
\begin{eqnarray}
\label{bk20}
Q_s(t) = a_1(t) q_s(0) + a_2(t) {\dot {q}_s(0)}
\, ,
\end{eqnarray}
and construct the monodromy matrix
\begin{eqnarray}
\label{bk21}
\left (
\begin{array}{c}
Q_s(t) \\
{\dot {Q}_s(t)}
\end{array}
\right ) = {\hat {M}}
\left (
\begin{array}{c}
q_s(0) \\
{\dot {q}_s(0)}
\end{array}
\right )\, ,
\end{eqnarray}
The determinant of the monodromy matrix (\ref{bk21}) equals to $W^2(t)$ and then following the
known procedure \cite{5}, \cite{5a} we end up with the generalized for discrete reservoir spectrum effective CL Hamiltonian 
\begin{eqnarray}
\label{bk22}
H_{eff}^d = W^2(t) \frac{1}{2}P_s^2 + W^{-2}(t) \frac{1}{2}Q_s^2 
\, .
\end{eqnarray}
The derivation of this effective Hamiltonian $H_{eff}^d$ is the main result of our paper
and it is ready for further inspection and various applications.
Unlike the standard CL Hamiltonian (\ref{bk2}) where with time going on
the oscillator $s$ irreversibly loses its energy and therefore the amplitude of the oscillations
decays exponentially, for our Hamiltonian (\ref{bk22}) this picture holds only for the initial $k=0$ cycle.
Then the oscillator $s$ not only transfers its energy to the discrete spectrum $R$ oscillators but as well 
receives the energy from the reservoir. We show time evolution of the oscillator $s$ amplitude
in the fig. 1.
In the cycle $k=0$ the initial $s$-oscillator state decays exponentially (the upper panel
in the fig. 1). This decay is accompanied by the monotonous growth of the $R$ oscillators
amplitudes (second and third from the top panels in the fig. 1). In the cycle $k=1$ the $s$
oscillator amplitude increases because of the reverse $R \to s$ transitions. 
Meanwhile the states of the
reservoir are depopulated-populated almost synchronously and it yields to the intense one-component Loschmidt echo
signal. In the next $k > 1$ cycles, the synchronisation of the reverse transition is gradually destroyed.
Then the Loschmidt echo signals become multi-component ones. Eventually, when the width of the echo signal
becomes comparable with the cycle period, one observes mixing phenomena and crossover to irregular, stochastic
like dynamics \cite{SI77}, \cite{AL05}.

{\bf{Evolution of quantum states of the $s$ oscillator.}}
${\, }$ 
To get the standard quantum mechanical coordinate - momentum commutation relation we have to define the
momentum $P_s$ conjugated to $Q_s$ (\ref{bk20}) as an operator
\begin{eqnarray}
\label{bk23}
P_s \equiv W(t){\dot {Q}_s} 
\, .
\end{eqnarray}
The same function $W(t)$ determines non-zero matrix-elements 
\begin{eqnarray}
\label{bk24}
(Q_s^2)_{\nu \, \nu -1} = \frac{\nu }{2}W^{2}(t)  
\, .
\end{eqnarray}
Therefore if the $s$ oscillator initial ($t=0$) state has the quantum number $\Lambda $,
the wave function $\Psi _\Lambda (Q_s,t)$ at the time $t$ is uniquely determined
by the eigen function $\Phi _\Lambda (Q_s)$ of the initial state
\begin{eqnarray}
\label{bk25}
\Psi _\Lambda (Q_s, t) = A_s(\Lambda , t) \Phi _\Lambda (Q_s)
\, ,
\end{eqnarray}
where the wave function amplitude $A_s(\Lambda , t) \equiv (W(t))^\Lambda $.
We conclude from the (\ref{bk25}) that the ground state wave function is time independent (as it should be),
whereas the wave functions for the excited states change in time manifesting recurrence cycle behavior.
For the initial cycle $k=0$
\begin{eqnarray}
\label{bk26}
A_s(\Lambda , t) = \exp (-\Lambda \eta t/2)
\, .
\end{eqnarray}
Because in the harmonic approximation only transitions where the quantum numbers are changed by $\pm 1$
are not forbidden, the global ($s$ oscillator and $R$ oscillators) wave function is
the superposition (with time dependent coefficients) of the harmonic oscillator wave functions with the total quantum
number $\Lambda $, i.e.,
\begin{eqnarray}
\label{bk27}
\Psi _\Lambda (t) = \sum _{\nu _s\, ;\{\nu _n\}} A(\nu _s\, ; \{\nu _n\}\, ; t) \Phi _{\nu _s}(q_s) \prod _{n} \Phi _{\nu _n}(q_n)
\, ,
\end{eqnarray}
where mentioned above selection rule requires that $\nu _s + \sum _{n}\nu _n = \Lambda $.
We show in the Fig. 2, the time evolution of the $\Lambda =4$ initial state of the $s$ oscillator.
For the excited state, the superposition (\ref{bk27}) includes the wave function of the ground state $(\Lambda _s\, \{0_n\})\, ;\,
((\Lambda -1)_s\, 1_n\, \{0_{n^\prime \neq n }\})\, ;\, ............$.
The coefficients $A$ entering the (\ref{bk27}) are determined by the product of the corresponding powers of the function $W(t)$.
The echo signal decreases with $\Lambda $, and as a result, the decay of the highly excited initial state becomes
irreversible (the upper panel in the fig. 2). Then less excited states also decay,
and in the next cycles, the excitation energy is redistributed (as one can expect approaching to the
equilibrium state) between low energy excited states of the $R$
oscillators (the lowest panel in the fig. 2).

{\bf{Conclusion.}}
${\, }$ 
To summarize, 
in this paper we investigated quantum
dynamics for a quantum oscillator $s$ coupled bilinearly to a reservoir $R$ 
with dense discrete spectrum of harmonic oscillators with close to $\omega _s$ frequencies. 
We show that for such systems dynamic evolution of any state of the CL Hamiltonian is reduced to recurrence cycles (oscillations)
due to reverse (from
the discrete spectrum  reservoir $R$ to $s$) transitions not existing in the limit of a macroscopic thermal bath. 
These reverse transitions lead to a partial restoration of the $s$ oscillator amplitudes (Loschmidt echo).
Width and number of the Loschmidt echo components increase with the recurrence cycle number $k$, eventually leading to
irregular, stochastic-like time evolution.
As a possible physical realization and evidence for the discussed above phenomena we
note recent experiments on ultra-fast (in $ps$ scale) energy exchange between high frequency vibrations \cite{TA11}
and observations of almost ballistic energy transport in molecular chains \cite{LK11}.
We anticipate also that considered above behavior of the CL Hamiltonian is relevant to describe experimental
data for radiationless transitions and ultrafast photo-chemical reactions.

\acknowledgements

E.K. acknowledges the financial support and hospitality of the Isaac Newton Institute for Mathematical Sciences,
and also FTP S \& SPPIR program.

\newpage

\centerline{Figure captions}

Fig. 1

Time evolution of the oscillator $s$ state with $\Lambda =1$, and for the $R$ oscillators
$n=1\, ,\, 3\, ,\, 4\, ,\, 7$ which were initially in the ground state.
The friction coefficient $\eta = 2.2$.

Fig. 2

Time evolution of the oscillator $s$ state with $\Lambda =4$,
and occupations of the following states $s$ and $R$ oscillators:
$(3_s\, 1_n)\, ,\, (2_s\, 1_n\, 1_{n^\prime }\, ,\, (1_s\, 1_n\, 1_{n^\prime}\, 1_{n^{\prime \prime }})\, ,\,
(1_n\, 1_{n^\prime}\, 1_{n^{\prime \prime }}\, 1_{n^{\prime \prime \prime }}))$.
The friction coefficient $\eta = 3.1$.


\begin{references}

\bibitem{SC94} 
R.Schinke, Photodissociation Dynamics, Cambridge Univ. Press., Cambridge (1994).


\bibitem{1} T.User, W.H.Miller, Phys. Repts., {\bf 199}, 73 (1991).


\bibitem{2}
S.Mukamel Principles of nonlinear optical spectroscopy, Oxford Univ. Press, Oxford (1995).

\bibitem{2a} D.M.Leitner, Adv. Chem. Phys., {\bf 130 b}, 205 (2005).

\bibitem{CL83} A.O.Caldeira, A.J.Leggett, Ann. Phys., {\bf 149}, 374 (1983); Physica A, {\bf 121}, 587 (1983).

\bibitem{LC87}
A.J.Leggett, S.Chakravarty, A.T.Dorsey, M.P.A.Fisher, A.Garg, M.Zweger, Rev. Mod. Phys., {\bf 59}, 1 (1987).

\bibitem{YS94} L.H. Yu, Ch.-P. Sun, Phys. Rev. A, {\bf 49}, 592 (1994).

\bibitem {ZW60} 
R.Zwanzig, Lectures in Theor. Phys., {\bf 3},
106 (1960).

\bibitem{BF07} 
V.A.Benderskii, L.A.Falkovsky, E.I.Kats, JETP Lett., {\bf 86}, 311 (2007).


\bibitem{BK08} V.A.Benderskii, E.I.Kats, JETP Lett., {\bf 88}, 387 (2008).

\bibitem{BK09} V.A.Benderskii, L.N.Gak, E.I.Kats, JETP, {\bf 135}, 176 (2009); ibidem {\bf 136}, 589 (2009).



\bibitem{BE53} 
H.Bateman, A.Erdelyi,
Higher Transcendental Functions, vol.2, McGraw Hill, New
York (1953).

\bibitem{5} 
R.P.Feynman, A.R.Hibbs, Quantum mechanics and path integrals, McGraw Hill, New York (1965).

\bibitem{5a}
H.Kleinert, Path integrals in quantum mechanic, statistics and polymer physics,
World Scientific, Singapore (1994).


\bibitem{SI77} 
Ya.G.Sinai, Introduction to Ergodic Theory, Princeton Univ. Press, Princeton (1977).


\bibitem{AL05}
J.F.Alves, S.Luzzatto, V.Pinheiro, Ann. Inst. H.Poincare, Anal. Non-Lineaire, {\bf 22}, 817 (2005).


\bibitem{TA11}
M.C.Thielges, J.Y.Axup, D.Wong, H.S.Lee, J.K.Chung, P.G.Schultz, M.D.Feyer,
J.Phys. Chem. B, {\bf 115}, 11294 (2011).

\bibitem{LK11}
Z.Lin, P.Keifer, I.V.Rubtsov, J.Phys. Chem. B, {\bf 115}, 5347 (2011).




\end{references}
\end{document}